\documentclass[aps,amsmath,amssymb,10pt,showpacs]{revtex4}
\usepackage{graphicx}
\usepackage{mathrsfs}
\usepackage{bbm}

\begin{document}

\title{Monsters, black holes and the statistical mechanics of gravity\footnote{This invited Brief Review for Modern Physics Letters A is based on several talks given by the authors in 2007--2009.}}
\author{Stephen~D.~H.~Hsu}\email{hsu@uoregon.edu}\affiliation{Institute of Theoretical Science, University of Oregon,
Eugene, OR 97403-5203, USA}
\author{David Reeb}\email{dreeb@uoregon.edu}\affiliation{Institute of Theoretical Science, University of Oregon,
Eugene, OR 97403-5203, USA}

\date{August 2009}

\begin{abstract}
We review the construction of {\it monsters} in classical general relativity. Monsters have finite ADM mass and surface area, but potentially unbounded entropy. From the curved space perspective they are objects with large proper volume that can be glued on to an asymptotically flat space. At no point is the curvature or energy density required to be large in Planck units, and quantum gravitational effects are, in the conventional effective field theory framework, small everywhere. Since they can have more entropy than a black hole of equal mass, monsters are problematic for certain interpretations of black hole entropy and the AdS/CFT duality.

In the second part of the paper we review recent developments in the foundations of statistical mechanics which make use of properties of high-dimensional (Hilbert) spaces. These results primarily depend on {\it kinematics} -- essentially, the geometry of Hilbert space -- and are relatively insensitive to dynamics. We discuss how this approach might be adopted as a basis for the statistical mechanics of gravity. Interestingly, monsters and other highly entropic configurations play an important role.
\end{abstract}


\maketitle

\section{Introduction: What is Entropy?}
Statistical (microcanonical) entropy $S$ is the logarithm of the number of distinct microstates $\psi$ of a system consistent with some imposed macroscopic properties, such as a restriction on the total energy. Thus, the entropy $S$ is proportional to the logarithm of the dimensionality of the Hilbert space of allowed $\psi$'s and measures the amount of information that is encoded in a particular microstate $\psi$. Unitarity forbids any change in the size of this Hilbert space during time evolution of the system, but entropy may increase if the macroscopic description changes so that more microstates become consistent with it. 

Without a theory of quantum gravity, we do not know, so cannot count, the microstates of black holes (for results in string theory, see \cite{StringTheoryEntropySV,StringTheoryEntropyMSW}). But it has been established \cite{HawkingEntropy} semiclassically that a large black hole of mass $M$ emits thermal radiation of temperature $T \sim M^{-1}$, so the entropy in this Hawking radiation is of order the area of the hole: $S = \int dQ / T \sim \int dM\,M \sim A$ (we use Planck units $\hbar = c = G = 1$ throughout). Strictly speaking, the Hawking process applies only to the semiclassical part of the evaporation, but the final quantum part releases at most of order the Planck energy, which can be made negligible compared to the initial mass of the hole and is thus unlikely to change the scaling with $M$ of the total amount of radiation entropy. A total black hole entropy of $S_{BH} = A/4$, corresponding to an entropy density $\sim10^{69}\,{\rm bit}/{\rm m}^2$ on the horizon, is consistent with other evidence ranging from black hole thermodynamics \cite{BekensteinEntropy,HawkingEntropy} to string theory \cite{StringTheoryEntropySV,StringTheoryEntropyMSW}, although there are other interpretations of this area entropy as well \cite{NatureOfBHEntropy}.

A black hole has much more entropy than ordinary matter configurations of the same size \cite{footnote1} and energy. For ordinary matter in flat space, the following bound \cite{th} applies: $S ~<~ A^{3/4}$. This result can be derived as follows. Given a thermal region of radius $R$ and temperature $T$, we have $S \sim T^3 R^3$ and $E \sim T^4 R^3$. Requiring $E < R$ (using the hoop conjecture -- a criterion for gravitational collapse \cite{hoop,bhEG,bhSDHH}) then implies $T < R^{-1/2}$ and $S < R^{3/2} \sim A^{3/4}$. The use of a temperature $T$ in this derivation is justified because the entropy of a system of fixed size and total energy is maximized in thermal equilibrium. 

In Planck units, and for macroscopic objects, the gap between $A$ and $A^{3/4}$ scaling is prodigious. Part of the motivation for the work described here was to understand whether this gap in scaling could be closed by considering curved, rather than flat, space. Another related question, also addressed below, is whether black holes are the most entropically dense objects in the universe. The answers to these questions are (at least in classical general relativity): Yes, non-black hole configurations can be found which have more than $A^{3/4}$ entropy, although such configurations are very non-Euclidean, and No, black holes are not the most entropic objects of fixed surface area and mass, unless some further principle (presumably of quantum nature) is introduced into the theory to remove even higher entropy configurations. 

The highly entropic objects we have found all collapse into black holes, which is problematic if black hole evaporation is unitary, since unitary evolution cannot map a larger Hilbert space into a smaller one. (Of course, it is also possible that black hole evaporation violates unitarity \cite{HawkingPureMixed,TopologyChange}.) We discuss this further below.

\section{Constructing Monsters}
We present two examples of classes of such highly entropic configurations $\Sigma_0$ (matter+gravity). In both examples, the curvature of space on $\Sigma_0$ makes the ADM mass (i.e., the energy a distant external observer sees and that determines the black hole area after collapse and hence the eventual Hawking radiation entropy) and the surface area of the configuration much smaller than would be suspected from the proper internal volume, to which the initial entropy $S_{\Sigma_0}$ is proportional. In the case of ``{\it monsters}'' (Sect.~\ref{monsterssubsection}), this effect can be ascribed to large negative binding energy \cite{Hsuzero} which almost cancels the proper mass to yield a relatively small ADM mass. In Sect.~\ref{frwsubsection}, the Kruskal--FRW example, the reason is the non-monotonic behavior of the radius $r$ of 2-spheres across the outer Einstein-Rosen bridge.

\begin{figure}[t]
\includegraphics[width=2.5in]{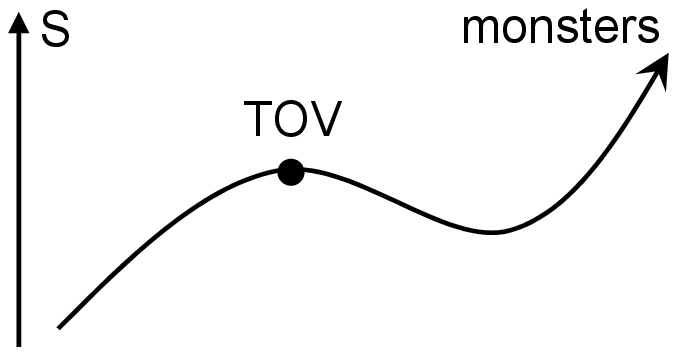}
\caption{As the 3-geometry inside a given 3-sphere $A$ is varied, it can accommodate different numbers $e^S$ of matter configurations. Stationary gravity-matter configurations (solutions to the Tolman-Oppenheimer-Volkoff equation) are local extrema of the entropy $S$, but, as one varies the internal 3-geometry, monster configurations can have unbounded entropy at fixed ADM mass $M$ and surface area $A$.\label{smfig}}
\end{figure}

Unlike ordinary configurations such as stars, galaxies, or even black holes, monster-like configurations have unbounded entropy at fixed ADM mass and surface area: Even if we force the spacetime to be asymptotically flat and fix its ADM mass at $M$ and if, moreover, we require all excited matter degrees of freedom to be contained within a 3-sphere of fixed surface area $A$ (this definition is unambiguous in the case of spherical symmetry, which our examples will obey), there are still an infinite number of matter+gravity configurations inside this surface which conform to these restrictions. In fact, imagine that, additionally, the 3-geometry (at some instant in time, e.g.~at a moment of time symmetry) inside the sphere is fixed and that one only looks for matter configurations which generate this given geometry (via the Einstein constraint equations of classical general relativity); then the entropy $S$ characterizing these matter configurations \emph{alone} is already unbounded as one varies the 3-geometry inside the surface $A$ (Fig.~\ref{smfig}). The stationary points of $S$ as a function of the 3-geometry correspond \cite{SorkinWaldZhang} to solutions of the Tolman-Oppenheimer-Volkoff equation (i.e., they are stationary stars, etc.), but for some interior 3-geometries the entropy $S$ can be much bigger and be even larger than $A$ or $M^2$ (typically, the configuration will not be stationary in this case). Sects.~\ref{monsterssubsection} and \ref{frwsubsection} describe examples of such configurations. Clearly, then, if the 3-geometry inside the surface is \emph{not} specified at all, one has to ascribe an infinite entropy to the system.

\subsection{Monsters}\label{monsterssubsection}
\begin{figure}[t]
\includegraphics[width=4.7in]{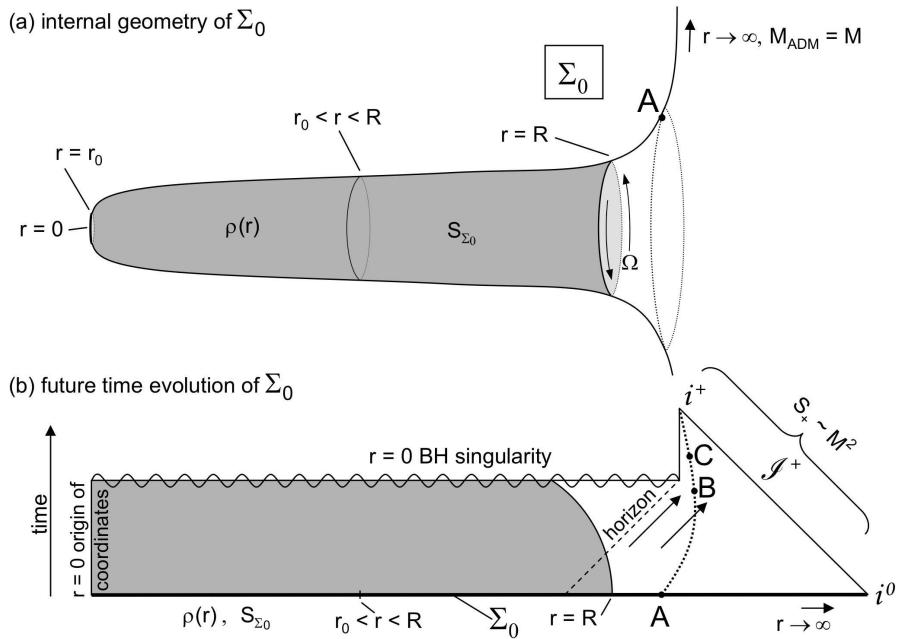}
\caption{(a) Embedding of the monster configuration $\Sigma_0$ into flat space with one angular dimension suppressed. The ``neck'' has proper length much bigger than $\left(R-r_0\right)$, due to the huge factor $\epsilon(r)^{-1/2}$, and contains all of the initial entropy $S_{\Sigma_0}$. For $r>R$ the geometry is just that of a Schwarzschild slice with mass $M = M_{ADM}$. (b) The monster's future time evolution is similar to ordinary gravitational collapse: (almost) all matter and entropy, if it was not already initially, will fall behind a horizon (infall of outer layers soon creates trapped surfaces) and form a black hole which then evaporates, radiating away entropy $S_+ \sim M^2 < S_{\Sigma_0}$ past the external observer to future infinity $\mathscr{I}^+ \cup i^+$.\label{monsterfig}}
\end{figure}

Our first example is a ball of material which is on the verge of collapsing to form a black hole. Its energy density profile is arranged to produce a curved internal space with large proper volume (see Fig.~\ref{monsterfig}(a)). The configuration is spherically symmetric, defined by initial data on a Cauchy slice $\Sigma_0$ at a moment of time symmetry (i.e., configuration initially ``at rest'') without (marginally) trapped surfaces, so that $\Sigma_0$ has geometry
\begin{equation}
\label{MetricSigma0}
ds^2 \bigr\vert_{\Sigma_0}
= \epsilon(r)^{-1} dr^2+r^2 d \Omega^2~,\quad K_{ab} \bigr\vert_{\Sigma_0} = 0\,,
\end{equation}
with $\epsilon(r) > 0$. For given initial matter distribution $\rho(r)$, Einstein's (constraint) equations determine
\begin{equation}
\epsilon(r)=1-\frac{2M(r)}{r}\,,
\end{equation}
where
\begin{equation}
M(r) = 4 \pi \int_0^r dr' \, {r'\,}^2 \rho(r')\,.
\end{equation}
If a configuration has radius $R$, i.e.~$\rho(r>R) = 0$, its ADM energy is $M = M(R)$. This quantity is constant during time evolution of the configuration (Birkhoff's theorem), and, if it collapses to a black hole, equals its mass.

Now, consider a semiclassical configuration (``monster'' \cite{MonsterPaper,SorkinWaldZhang}, Fig.~\ref{monsterfig}(a)) with radius $R \gg 1$ that yields
\begin{equation}
\epsilon(r)=\left(\frac{r_0}{r}\right)^{\gamma}~,\quad\quad r_0 < r < R\,,
\end{equation}
with some $\gamma>0$ and $r_0 \ll R$ (to avoid poles), so that the configuration comes increasingly closer to forming trapped surfaces as $r \nearrow R$ (long ``neck'' in Fig.~\ref{monsterfig}(a)). It has ADM mass
\begin{equation}
M = \frac{R}{2}\left(1-\epsilon(R)\right) \approx \frac{R}{2} \sim R
\end{equation}
and energy density
\begin{equation}
\rho(r)=\frac{M'(r)}{4\pi r^2}
\approx \frac{1}{8\pi r^2} \sim \frac{1}{r^2}~,\quad\quad r_0 < r < R\,.
\end{equation}
Finally, with a relation $s = \alpha \rho^{\beta} \sim \rho^{\beta}$ between energy and entropy density of the matter ($\alpha = {\cal O}(1)$), the initial entropy is
\begin{equation}
S_{\Sigma_0} = 4 \pi \int_0^R dr \, r^2 \epsilon(r)^{-1/2} s(r)  \sim  \frac{R^{3-2\beta+\gamma/2}}{r_0^{\,\gamma/2}} \sim A^{3/2-\beta+\gamma/4}\,,
\end{equation}
with the area $A \sim M^2$ of the black hole formed in collapse of this monster.

It is now evident that, if $\beta$ is constant, one can always find configuration parameters $\gamma$ such that the entropy of the monster exceeds area scaling (hence, the name). This is the case, e.g., if we model the matter (initially) as a perfect fluid with equation-of-state parameter $w$. Then $\beta = 1/\left(1+w\right)$, and we would just have to choose $\gamma > 1$ for a photon gas ($w = 1/3$) or $\gamma > 2$ for dust ($w = 0$; we assume the dust particles carry some kind of label or have spin).

Fig.~\ref{monsterfig}(b) depicts the time evolution of a monster, which resembles ordinary gravitational collapse. The main difference is that, due to our construction, the entropy $S_{\Sigma_0}$ on the initial Cauchy slice can be much bigger than the entropy $S_+=A/4$ on future infinity, assuming that black hole evaporation is unitary and the standard assumptions about Hawking radiation hold. In order to preserve unitarity (or the AdS/CFT duality \cite{AdSCFTReview,Freivogel:2005qh}) one would somehow have to excise monsters with $S > A$ from the Hilbert space. Monsters with sufficiently high entropy are therefore semiclassical configurations with \emph{no} corresponding microstates in a quantum theory of gravity. 

Note, if $r_0$ is chosen a few orders of magnitude above the Planck length, all involved densities $\rho(r)$ and $s(r)$ are sub-Planckian, so that our semiclassical analysis naively applies. Furthermore, Bousso's covariant entropy bound \cite{boussobound} holds in the semiclassical monster spacetime since it falls under the general class of spacetimes for which a general theorem \cite{FMW} applies (this assumes no large entropy gradients due to, e.g., shockwaves during evolution, which seems plausible, but has not been proven).

\subsection{Kruskal--FRW gluing}\label{frwsubsection}
\begin{figure}[t]
\includegraphics[width=4.7in]{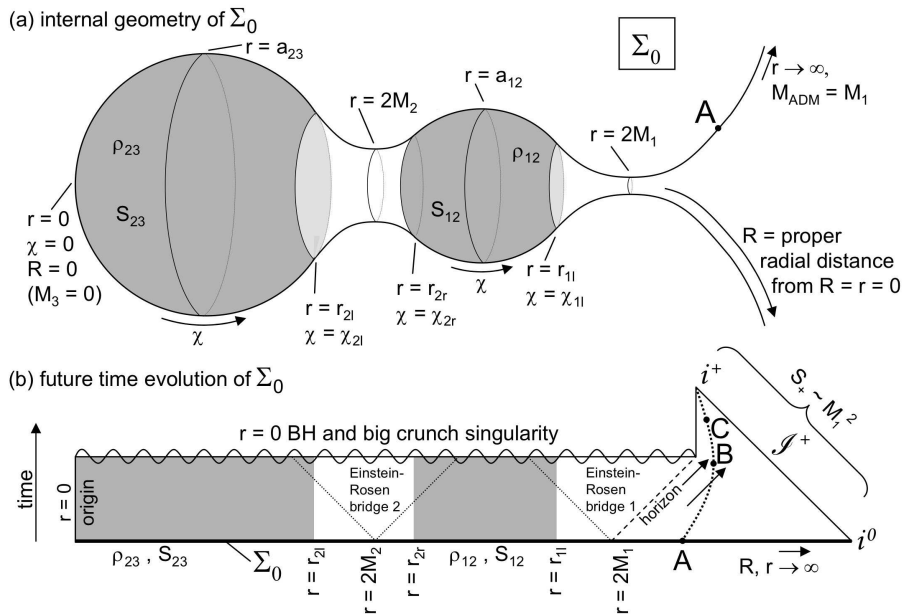}
\caption{(a) Embedding of a glued Kruskal-FRW initial slice $\Sigma_0$ into flat space with one angular dimension suppressed. $R$ is the proper radial distance from the innermost point and $r=r(R)$ gives the radius of the 2-sphere labeled $R$. Additional or larger closed FRW pieces could be adjoined, and there could also be a second asymptotic Kruskal piece (even with mass parameter different from $M_1$) if the far left were not closed off with a 3-sphere. (b) By considering the rightmost Einstein-Rosen bridge, standard energy conditions suffice to show that a singularity will form and that the external observer will see a black hole of mass $M_1$ whose Hawking radiation then contains potentially much less entropy $S_+\sim M_1^{\,2}$ than was present on $\Sigma_0$. In the case of pressureless dust, the time evolved spacetime can be given analytically as Kruskal spacetimes and FRW universes appropriately sewn together (Oppenheimer-Snyder collapse).\label{frwfig}}
\end{figure}

The second example \cite{UnitarityPaper} consists of slices of closed FRW universes which are glued together across Einstein-Rosen bridges, eventually connecting to a large asymptotically flat universe (Fig.~\ref{frwfig}(a)). Again, a larger proper volume can be accommodated at fixed ADM mass. The configuration is specified, as before, by initial data on a spherically symmetric and time symmetric ($K_{ab} \vert_{\Sigma_0} = 0$) Cauchy slice $\Sigma_0$: we take the part of a constant-time slice of the Kruskal spacetime with mass $M_1$ (e.g., part of the $U+V = 0$ slice, in usual Kruskal coordinates) that contains one asymptotic region with outside observer A, the Einstein-Rosen bridge at its maximal extent $r=2M_1$ and the piece $r_{1{\rm l}} > r > 2M_1$ of the other asymptotic region (right part in Fig.~\ref{frwfig}(a)). This is then glued onto the part $\chi < \chi_{1{\rm l}}$ of the hypersurface $ds^2 = a_{12}^{\,2} \left( d\chi^2+\sin^2 \chi \, d\Omega^2 \right)$ representing a closed FRW universe at the instant of its maximal expansion $a_{12}$. By cutting this 3-sphere off at $\chi = \chi_{2{\rm r}}$, a second piece of Kruskal containing an Einstein-Rosen bridge can be joined, etc. In our notation the integer subscript $n$ denotes the $n$-th Einstein-Rosen bridge, and l (r) denote left (right), see Fig.~\ref{frwfig}.

Matching the geometry across the common boundary requires the transverse metric to be continuous and continuously differentiable (i.e.~the extrinsic curvature $K^{(3)}_{ab}$ has to be the same on either side); its second derivative can be discontinuous, as is the energy density $\rho$, consistent with Einstein's equation $G_{ab} = 8\pi T_{ab}$. At the rightmost joining surface in Fig.~\ref{frwfig}(a), continuity of the transverse metric means equality of the areas of the spherical sections $\chi=\chi_{1{\rm l}}$ and $r=r_{1{\rm l}}$, i.e.
\begin{equation}
\label{equalA}
a_{12}\sin \chi_{1{\rm l}} = r_{1{\rm l}}\,.
\end{equation}
And equality of extrinsic curvatures is, in the case of spherical symmetry, equivalent to continuous differentiability of the area $A(R)$ of 2-spheres with respect to proper radial distance $R$:
\begin{equation}
\frac{d}{a_{12}\,d\chi} \left( 4\pi a_{12}^{\,2} \sin^2 \chi \right) \biggr\vert_{\chi=\chi_{1{\rm l}}} \,=\, 
\frac{d}{- \left( 1-2M_1/r \right)^{-1/2}\,dr} \left( 4\pi r^2 \right) \biggr\vert_{r=r_{1{\rm l}}}\,,
\end{equation}
which forces $\chi_{1{\rm l}} \in [\pi/2,\pi)$ and, with (\ref{equalA}),
\begin{equation}
\label{diffA}
2M_1 = a_{12}\sin^3\chi_{1{\rm l}}\,.
\end{equation}
Equations like (\ref{equalA}) and (\ref{diffA}) hold at every joining surface, with a modified constraint $\chi_{{\rm r}} \in [0,\pi/2]$ if joining just \emph{right} of an Einstein-Rosen bridge. From these formulae, a configuration like Fig.~\ref{frwfig}(a) can be constructed, e.g., in the following way: first pick masses $M_1,~M_2, \ldots$ describing the Kruskal pieces ($M=0$ forces the construction to an end), then sizes $a_{12},~a_{23} \ldots$ of the FRW pieces subject to constraints $a_{12} \ge 2M_1,~2M_2$, etc. $\Sigma_0$ is then uniquely determined.

Invoking Friedmann's equation with vanishing instantaneous expansion, the FRW pieces have energy density $\rho_{12}=3/8\pi a_{12}^{\,2}$. With $s \sim \rho^{\beta}$, the entropy of one piece becomes
\begin{equation}
\label{S12equation}
S_{12} = 4 \pi a_{12}^{\,3}s \int_{\chi_{2{\rm r}}}^{\chi_{1{\rm l}}} d\chi \, \sin^2\chi \, \sim\,
a_{12}^{\,3-2\beta} \left[ \chi_{1{\rm l}}-\chi_{2{\rm r}}-\frac{1}{2}\sin 2\chi_{1{\rm l}} +\frac{1}{2}\sin 2\chi_{2{\rm r}} \right]\,.
\end{equation}
The bracket in (\ref{S12equation}) approaches $\pi = {\cal O}\left(1\right)$ as $a_{12}$ becomes a few times bigger than $2M_1$ and $2M_2$. In that case, the total entropy on $\Sigma_0$ is
\begin{equation}
\label{initialSFRW}
S_{\Sigma_0} =
S_{12}+S_{23}+\ldots \,\sim\,
a_{12}^{\,3-2\beta} + a_{23}^{\,3-2\beta}+\ldots~,
\end{equation}
and so can be made arbitrarily big (for any $\beta=1/\left(1+w\right)<3/2$) by either taking the size of the FRW pieces or their number to be large.

Evolved forward in time (Fig.~\ref{frwfig}(b)), the entropy in the Hawking radiation that passes the external observer and reaches future infinity is $S_+ \sim M_1^{\,2}$, so again is potentially much less than the entropy on the initial slice (\ref{initialSFRW}). As in the case of monsters, the Kruskal-FRW configurations are reasonable semiclassical initial data insofar as all involved densities are well sub-Planckian (if the FRW pieces are a few orders of magnitude bigger than the Planck length).  The spacetimes do not violate the covariant entropy bound by the same arguments \cite{FMW} as before (cf.~also \cite{boussobound} for more specific discussion of entropy bounds in closed FRW universes).

\section{Evolution and Singularities}
\begin{figure}[t]
\includegraphics[width=2.5in]{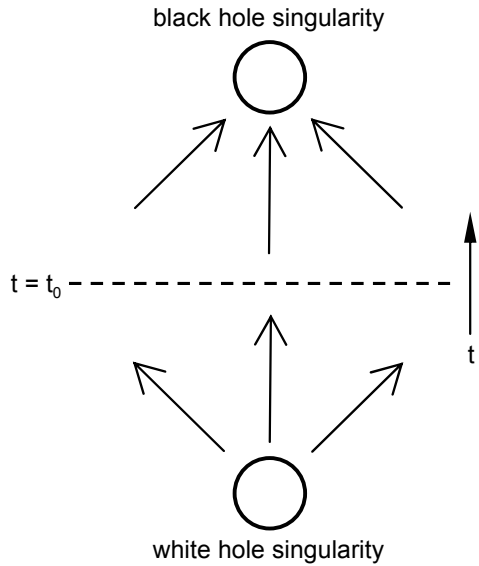}
\caption{An isolated monster (time-symmetric configuration at $t = t_0$) evolved forward in time becomes a black hole with a future singularity. The same monster therefore emerges from a past white hole singularity.\label{whiteholefig}}
\end{figure}

Both types of configurations have the pathological property that, under isolated evolution, they must have emerged from a past singularity (white hole; see Fig.~\ref{whiteholefig}). This can be seen via backward evolution of the time-symmetric initial data, noting that forward evolution leads to a black hole and future singularity. The monster itself can be thought of as an object whose negative gravitational binding energy almost cancels the positive kinetic and rest mass energy of its constituents. In Fig.~\ref{whiteholefig}, the monster explodes out of an initial white hole singularity. Because of the large gravitational binding energy, the constituents are unable to separate to infinity, but rather reach a turning point at $t = t_0$ and subsequently collapse back into a black hole.

To avoid the white hole singularity, one can relax the assumption of isolation, and consider monster initial data at $t = t_0$, perhaps constructed ``in the laboratory'' by outside intervention. One can show that the configurations with $S > A$ cannot be constructed, even via intervention by an arbitrarily advanced civilization \cite{SorkinWaldZhang,MonsterPaper}; that is, there seem to be fundamental physical limits on the construction of monsters. Despite their pathologies, these configurations represent valid semiclassical states of a matter-gravity system: they are all locally well behaved, in particular do not require large energy or entropy densities, and -- if present in the Hilbert space -- could be accessible via tunneling starting from an ordinary matter configuration with the same quantum numbers (ADM energy, angular momentum, charge).

\section{Quantum Foundations of Statistical Mechanics}\label{wintersection}
Recently, the foundations of statistical mechanics have been established as a consequence of the geometry of high-dimensional Hilbert spaces \cite{Winters1,Gemmer}. 

Consider a large system subject to a linear constraint $R$ (e.g., that it be in a superposition of energy eigenstates with the energy eigenvalues all being below some $E_{max}$), which reduces its Hilbert space from ${\cal H}$ to a subspace ${\cal H}_R$. Divide the system into a subsystem $X$, to be measured, and the remaining degrees of freedom which constitute an environment $E$, so ${\cal H} = {\cal H}_X \otimes {\cal H}_E$ and 
 \begin{equation}
 \rho_X \equiv\rho_X(\psi)= {\rm Tr}_E \vert \psi \rangle \langle \psi \vert
\end{equation}
is the density matrix which governs measurements on $X$ for a given pure state $\psi$ of the whole system. Note the assumption that these measurements are local to $X$, hence the trace over $E$.

It can be shown \cite{Winters1}, using the concentration of measure on hyperspheres \cite{HDG} (Levy's theorem), that for almost all $\psi \in {\cal H}_R$, 
\begin{equation}
\rho_X(\psi) \approx {\rm Tr}_E\left(\rho_*\right) \equiv \Omega_X\,,
\end{equation}
where $\rho_* = \mathbbm{1}_R / d_R$ is the equiprobable maximally mixed state on the restricted Hilbert space ${\cal H}_R$ ($\mathbbm{1}_R$ is the identity projection on ${\cal H}_R$ and $d_R$ the dimensionality of ${\cal H}_R$). $\Omega_X = {\rm Tr}_E\left(\rho_*\right)$ is the corresponding canonical state of the subsystem $X$. The result holds as long as $d_E \gg d_X$, where $d_{E,X}$ are the dimensionalities of  the ${\cal H}_E$ and ${\cal H}_X$ Hilbert spaces. (Recall that these dimensionalities grow exponentially with the number of degrees of freedom. The Hilbert space of an $n$ qubit system is $2^n$ dimensional.) In the case of an energy constraint $R$, $\Omega_X$ describes a perfectly thermalized subsystem with temperature determined by the total energy of the system.

To state the theorem in \cite{Winters1} more precisely, the (measurement-theoretic) notion of the \emph{trace-norm} is required, which can be used to characterize the distance between two mixed states $\rho_X$ and $\Omega_X$:
\begin{equation}
\Vert\rho_X-\Omega_X\Vert_1\equiv{\rm Tr}\sqrt{\left(\rho_X-\Omega_X\right)^2}\,.
\end{equation}
This sensibly quantifies how easily the two states can be distinguished by measurements, according to the identity
\begin{equation}
\label{tracesup}
\Vert\rho_X-\Omega_X\Vert_1 = {\rm sup}_{\Vert O\Vert\leq1}\,{\rm Tr}\left(\rho_XO-\Omega_XO\right)\,,
\end{equation}
where the supremum runs over all observables $O$ with operator norm $\Vert O\Vert$ smaller than 1 (projectors $P=O$ are in some sense the best observables, all other observables can be composed out of them, and they have $\Vert P\Vert=1$). Note that the trace on the right-hand side of (\ref{tracesup}) is the difference of the observable averages $\langle O\rangle$ evaluated on the two states $\rho_X$ and $\Omega_X$, and therefore specifies the experimental accuracy necessary to distinguish these states in measurements of $O$. The theorem then states that the probability that
\begin{equation}
\label{w1}
\Vert\rho_X\left(\psi\right)-{\rm Tr}_E\left(\rho_*\right)\Vert_1 ~\geq~ \epsilon+\sqrt{\frac{d_X^{\,2}}{d_R}}
\end{equation}
is less than $2\exp ( -\epsilon^2d_R/18\pi^3 )$. In words: let $\psi$ be chosen randomly (according to the Haar measure on the Hilbert space) out of the space of allowed states ${\cal H}_R$; 
the probability that a measurement on the subsystem $X$ \emph{only}, with measurement accuracy given by the rhs of (\ref{w1}), will be able to tell the pure state $\psi$ (of the entire system) apart from the maximally mixed state $\rho_*$ is exponentially small in the dimension of the space ${\cal H}_R$ of allowed states. Conversely, for almost all pure states $\psi$ any small subsystem $X$ will be found to be extremely close to perfectly thermalized (assuming the constraint $R$ on the whole system was an energy constraint).

As mentioned, the overwhelming dominance of ``typical'' states $\psi$ is due to the geometry of high-dimensional Hilbert space and the resulting concentration of measure. It is a consequence of kinematics only -- no assumptions have been made about the dynamics. Almost any dynamics -- i.e., choice of Hamiltonian and resulting unitary evolution of $\psi$ -- leads to the system spending nearly all of its time in typical states for which the density matrix describing any small subsystem $X$ is nearly thermal \cite{Winters2}. Typical states $\psi$ are maximally entangled, and the approach to equilibrium can be thought of in terms of the spread of entanglement, as opposed to the more familiar non-equilibrium kinetic equations. 

We can restate these results in terms of the entanglement entropy of the subsystem $X$, thereby making contact with the Second Law of Thermodynamics. The entanglement entropy is simply the von Neumann entropy of $\rho_X$:
\begin{equation}
\label{entanglemententropy}
S (X) = - {\rm Tr} \, \rho_X \log \rho_X\,.
\end{equation}
Using the same results on the concentration of measure, it can be shown \cite{Hayden} that, for the overwhelming majority of pure states $\psi$, $S(X)$ is extremely close to its maximum value $\log d_X$:
\begin{equation}
{\rm Prob} \left[ ~S(X) ~<~ \log d_X - \alpha - \beta ~ \right] 
\leq \exp \left( -  \frac{(d_X d_E - 1) C \alpha^2}{(\log d_X)^2} \right)\,,
\end{equation}
where $\beta = \frac{1}{\ln 2} \frac{d_X}{d_E}$ and $C = (8 \pi^2 \ln 2)^{-1}$. This implies that, for almost any choice of dynamics \cite{Winters2}, a subsystem $X$ is overwhelmingly likely to be found with nearly maximal entropy $S(X)$. The Second Law is seen to hold, in a probabilistic sense, even though the underlying dynamics is time-reversal invariant: density matrices $\rho_X$ with small entropy are highly improbable, and if $X$ is found in a low-entropy state, the entropy is more likely to increase than decrease over any macroscopic time interval. 

In our earlier discussion of monsters, the entropy we used was not the entanglement entropy $S(X)$ in (\ref{entanglemententropy}). Instead, we defined the entropy of a monster or black hole to be the logarithm of the number of distinct quantum states consistent with the imposed macroscopic conditions (e.g., fixed ADM mass $M$, object of area $A$). This entropy is directly proportional to the logarithm of the dimensionality of the Hilbert space consistent with the macroscopic description, so in the current discussion it is simply $\log d_X$ if we consider only the subset of $X$ configurations which are consistent with the description. Note that $\log d_X \geq S(X)$ and that, for typical pure states of the larger system, any subsystem $X$ will have entanglement entropy $S(X)$ near its maximum value $\log d_X$. Thus, within the framework for statistical mechanics discussed in this section, the entropy we defined earlier can be used to characterize the most likely (``equilibrium'') configurations to be found in $X$.

\section{Statistical Mechanics of Gravity?}
Can the quantum mechanical derivation of statistical mechanics given above be applied to gravity? For example, can we deduce the Second Law of Thermodynamics on semiclassical spacetimes (i.e., including, for example, large black holes)? 

This might seem overly ambitious since we currently lack a theory of quantum gravity. However, the results described above are primarily a consequence of the high-dimensional character of Hilbert spaces. If the state space of quantum gravity continues to be described by something like a Hilbert space, then its dimensionality will almost certainly be large, even for systems of modest size. Further, it seems a less formidable task to characterize some aspects of the state space of quantum gravity than to fully understand its dynamics. Indeed, for our purposes here we only consider semiclassical spacetimes.

Early attempts at quantization, culminating in the Wheeler-DeWitt equation, were based on the classical Hamiltonian formulation of general relativity \cite{WDW1,WDW2}. These led to a configuration space (``superspace'') of 3-geometries, modulo diffeomorphisms, and to the wavefunction, $\Psi [ h_{ab}, \phi ]$, of the universe as a functional over 3-metrics $h_{ab}$ and matter fields $\phi$. This description of the state space seems quite plausible, at least in a coarse grained sense, even if the fundamental objects of the underlying theory are something else (strings, loops, etc.). Let us assume that some form of short-distance regulator is in place (or, alternatively, that the dynamics itself generates such a regulator in the form of a minimum spacetime interval), so that we can neglect ultraviolet divergences.

Now consider the set of asymptotically flat, non-compact 3-geometries. Impose conditions on the asymptotic behavior so that the total ADM mass of the system is $M$, and further assume that all the energy density is confined to a region of surface area $A$. This results in a restricted state space ${\cal H}_R$. If the concentration of measure results apply to ${\cal H}_R$, then the observed properties of any small subsystem $X$ are likely to be the same as if the universe were in the equiprobable, maximally mixed state $\rho_* = \mathbbm{1}_R / d_R$. In the flat space case this leads to the usual canonical (Boltzmann) distribution in $X$.

However, from our monster analysis we know that we are already in trouble. Despite the short distance regulator and the restrictions on total energy and surface area, the Hilbert space dimension $d_R$ and entropy are infinite because of monsters and related configurations, see Fig.~\ref{smfig}. (In a sense this is a trivial consequence of the fact that they can have infinite proper volume but nevertheless be glued into the region of interest with surface area $A$.) Without a {\it further regularization} which limits the proper volumes and entropies of monster-like configurations, the maximally mixed state is ill-defined and we cannot recover the familiar thermodynamics of semiclassical spacetimes in the same way as in Sect.~\ref{wintersection} for ordinary quantum systems. In effect, to obtain any reasonable results we have to eliminate the highest entropy configurations from the state space \cite{footnote2}.
 
For this approach to produce the familiar results from ordinary and black hole thermodynamics, it is therefore necessary to invoke some new principle which excises the $S > A/4$ monsters from the state space. (Indeed, as discussed earlier, such an excision was already suggested by the requirement that black hole evolution be unitary, although it is {\it not} required by the covariant entropy bound \cite{boussobound}.) Once this is done, Schwarzschild black holes become the most highly entropic objects of mass $M$ and $A = 16 \pi M^2$.
It then seems possible that the statistical mechanics of gravitational systems could result from typicality of the state $\Psi [ h_{ab}, \phi ]$. In particular, one might be able to deduce a modification of the Second Law into a Generalized Second Law that takes into account the entropy of black holes and of other curved space objects.

\section{Conclusions}
Classical general relativity allows configurations of fixed ADM mass and surface area, but unbounded entropy (``monsters''). These configurations can be constructed as initial data such that at no point are energy or entropy densities, or curvatures, large in Planck units. Thus, under the usual assumptions about gravity as an effective field theory, they are well described in the semiclassical approximation.

It is of course not known whether such configurations persist in the quantum theory of gravity. If they do, their existence seems problematic for unitary evaporation of black holes \cite{UnitarityPaper} and for the AdS/CFT correspondence \cite{Freivogel:2005qh}. If, to the contrary, they are to be excised from the theory, some new fundamental principle is required.

In the second part of this review we studied a fundamentally quantum approach to statistical mechanics. The high dimensionality of Hilbert space and consequent concentration of measure are used to show that almost any pure state will lead to approximate canonical behavior of the density matrix of small subsystems. This approach also provides a probabilistic justification of the Second Law of Thermodynamics. We investigated whether a similar framework can be applied to gravitational systems. The existence (or non-existence) of monster-like states plays a central role in the outcome: we conclude that this approach cannot work in the presence of gravity unless monster-like states are indeed excised from the theory.

\section*{Acknowledgments}
The authors thank Sean Carroll, Nick Evans, Sabine Hossenfelder, Ted Jacobson, Hirosi Ooguri, John Preskill, Lee Smolin, Rafael Sorkin and Mark Wise for useful discussions. The authors are supported by the Department of Energy under DE-FG02-96ER40969.


\begin{thebibliography}{00}
\bibitem{footnote1}
Note, we need to restrict the size of the object as well as its total energy. An object with fixed total energy $E = T^4 R^3$, but no restriction on $R$, can have infinite entropy: we can take $R\rightarrow\infty$ and $T\rightarrow 0$ with $E$ fixed, so that $S = T^3 R^3 = E / T \rightarrow \infty$.
\bibitem{footnote2}
Of course, it is also possible that the initial pure state is atypical and subsequent dynamics somehow keeps the state in a very atypical region of the Hilbert space over very long time scales, so that the highly entropic configurations are essentially never sampled. In that case one cannot deduce the thermodynamic properties of the system from the concentration of measure phenomenon (i.e., typicality) alone: the system does not actually reach ultimate equilibrium.
\end{thebibliography}
\end{document}